\documentclass[conference]{IEEEtran}
\IEEEoverridecommandlockouts
\usepackage{cite}
\usepackage{amsmath,amssymb,amsfonts}
\usepackage{algorithmic}
\usepackage{graphicx}
\usepackage{textcomp}
\usepackage{xcolor}
\usepackage{siunitx}
\usepackage{hyperref}

\newcommand{\mult}{\nonscript\:}

\begin{document}

\title{Signal Restoration and Channel Estimation\\for Channel Sounding with SDRs\\
\thanks{This work has been supported by the Federal Ministry of Education and Research of the Federal Republic of Germany as part of the FunKI project with grant number 16KIS1182.}
}

\author{%
    \IEEEauthorblockN{%
      Julian Ahrens\IEEEauthorrefmark{1},
      Lia Ahrens\IEEEauthorrefmark{1},
      Michael Zentarra\IEEEauthorrefmark{1} and
      Hans D.\ Schotten\IEEEauthorrefmark{1}\textsuperscript{,}\IEEEauthorrefmark{2}}%
    \IEEEauthorblockA{%
      \begin{tabular}{c}%
        \IEEEauthorrefmark{1}%
        \textit{Intelligent Networks}\\%
        \textit{German Research Center for Artificial Intelligence}\\%
        Kaiserslautern, Germany%
      \end{tabular}%
      \qquad%
      \begin{tabular}{c}%
        \IEEEauthorrefmark{2}%
        \textit{Institute for Wireless Communication and Navigation}\\%
        \textit{Technische Universit\"{a}t Kaiserslautern}\\%
        Kaiserslautern, Germany%
      \end{tabular}%
      \\%
      \{julian.ahrens,lia.ahrens,michael.zentarra,schotten\}@dfki.de}}

\maketitle

\begin{abstract}
In this paper, the task of channel sounding using software defined radios (SDRs) is considered.
In contrast to classical channel sounding equipment, SDRs are general purpose devices and require additional steps to be implemented when employed for this task.
On top of this, SDRs may exhibit quirks causing signal artefacts that obstruct the effective collection of channel estimation data.
Based on these considerations, in this work, a practical algorithm is devised to compensate for the drawbacks of using SDRs for channel sounding encountered in a concrete setup.
The proposed approach utilises concepts from time series and Fourier analysis and comprises a signal restoration routine for mitigating artefacts within the recorded signals and an encompassing channel sounding process.
The efficacy of the algorithm is evaluated on real measurements generated within the given setup.
The empirical results show that the proposed method is able to counteract the shortcomings of the equipment and deliver reasonable channel estimates.
\end{abstract}


\section{Introduction}
In modern mobile communications systems, robust estimates of the core characteristics of all relevant communications channels are essential to achieving the targeted performance and reliability.
To this end, extensive measurement campaigns have been undertaken~\cite{WBS18} to acquire data for appropriate channel modelling on the one hand and fitting parameters of existing standard channel models on the other hand.
The insights gathered during these investigations provided the foundation for the standardisation of the key parameters currently employed in mobile communications systems.
However, since these systems need to operate consistently across a variety of deployment environments, the parameters are often chosen in a very conservative manner, which decreases spectral efficiency of the system as a whole and generally leads to suboptimal performance.
With the emergence of specialised use cases and deployment scenarios in environments that are difficult to characterise in a generic manner (e.g.\ pico- and femtocells in industrial environments), deployment-specific tuning of system parameters provides a promising method for optimising system efficiency and performance.

Traditionally, the hardware employed to acquire the needed channel measurements is expensive and bulky to deploy~\cite{KMH08}, which makes analysis and tuning for specific deployments costly and cumbersome.
In contrast, software defined radios (SDRs) provide a more accessible and flexible alternative for channel sounding.
Furthermore, the required hardware is highly programmable and may be repurposed for other communication-related tasks once all required channel characterisations have been performed.
One disadvantage of using SDRs instead of traditional channel sounding equipment, however, is that their specifications are usually not as high~\cite{B210} and there is more uncertainty about the nature of possible inaccuracies or disturbances.
This may show up in the form of artefacts in the recorded signal, which in turn may lead to misestimation of channel properties.

In this work, a channel sounding setup based on two SDRs is considered.
Therein, the transmitting unit is of particularly high mobility, as it can be driven from a single laptop computer on battery power.
As expected and mentioned above, the setup produces artefacts in the recorded signals which may contaminate the channel estimates.
The main contribution of this work is a practical signal restoration algorithm that greatly reduces these measurement artefacts on the one hand and an encompassing channel sounding process that produces reasonable channel estimates on the other hand.

The remainder of the paper is structured as follows:
The hardware setup and the parameters governing the test transmissions are introduced in Section~\ref{sec:setting}.
Section~\ref{sec:effects} provides an analysis of different types of artefacts encountered during these transmissions and the subsequent channel estimation.
In Section~\ref{sec:restore}, the proposed procedures for signal restoration and channel estimation are presented.
The effects of the proposed algorithms are evaluated in Section~\ref{sec:results}.
Section~\ref{sec:conclusion} concludes the paper.

\section{Measurement setup}\label{sec:setting}
For the generation and measurement of the required signals, two different SDRs from the Ettus Research Universal Software Radio Peripheral (USRP) series are employed.
The USRPs are connected to their respective host PCs, which run GNU Radio for flexible signal generation and processing.

On the transmitter side, a USRP B210 is deployed.
A USB connection provides power and the data to be transmitted from its host PC.
The B210 has a maximum sampling rate of~\SI{61.44}{MS/s}~\cite{B210}.
The host PC generates the respective discrete-time signals digitally and sends them to the B210 for digital-to-analog conversion, filtering, power amplification, and IQ modulation~\cite{B210schematics}.

A USRP N310 operates as the receiver.
The N310 filters, amplifies, IQ demodulates, and analog-to-digital converts the received signal.
The digital signal is then sent from the N310 to the host PC via Gigabit Ethernet, where the complex signal is saved to a file for further processing.
The N310 provides sampling rates up to~\SI{153.6}{MS/s}~\cite{N310}.
However, the limiting factor of the setup is the Gigabit Ethernet connection.
Therefore, the sampling rate for both the signal generation on the transmitter side and the signal reception on the receiver side is set to \SI{25.6}{MS/s}.
The carrier frequency for all measurements is~\SI{2.48}{GHz}.

For the purpose of evaluation, two types of transmissions are performed: cable-bound and wireless.
For the cable-bound transmissions, the B210 output is connected via a coaxial cable and a~\SI{30}{dB} attenuator to the input of the N310, which records the signal.
For the wireless transmissions, both the transmitter and receiver are equipped with a VERT2450 omni-directional~\SI{2}{dBi} dual band antenna for operating frequencies 2.4-\SI{2.5}{GHz} and 4.9-\SI{5.9}{GHz}.
Here, two different configurations are implemented: a short-range line-of-sight configuration for calibration measurements and a longer-range non-line-of-sight configuration to capture the effects of a real-world channel.
The line-of-sight transmissions take place in a single room where the transmitter and receiver are spaced approximately~\SI{2.5}{m} apart.
For the non-line-of-sight transmissions, the N310 is placed outside a window in the second floor of a residential building.
The B210 is then moved across a parking lot and down into a street so that the line of sight is blocked by multi-level buildings.
The distance between the transmitter and receiver in this scenario is approximately~\SI{100}{m}.

In all experiments, the transmitter repeats the test signal indefinitely.
The receiver captures a pre-determined number of samples such that multiple periods of the transmitted signal are always included in the recording.

\section{Analysis of artefacts}\label{sec:effects}

\begin{figure}[tbp]
  \centering
  \includegraphics[trim=0 2 0 8,width=.75\linewidth,clip]{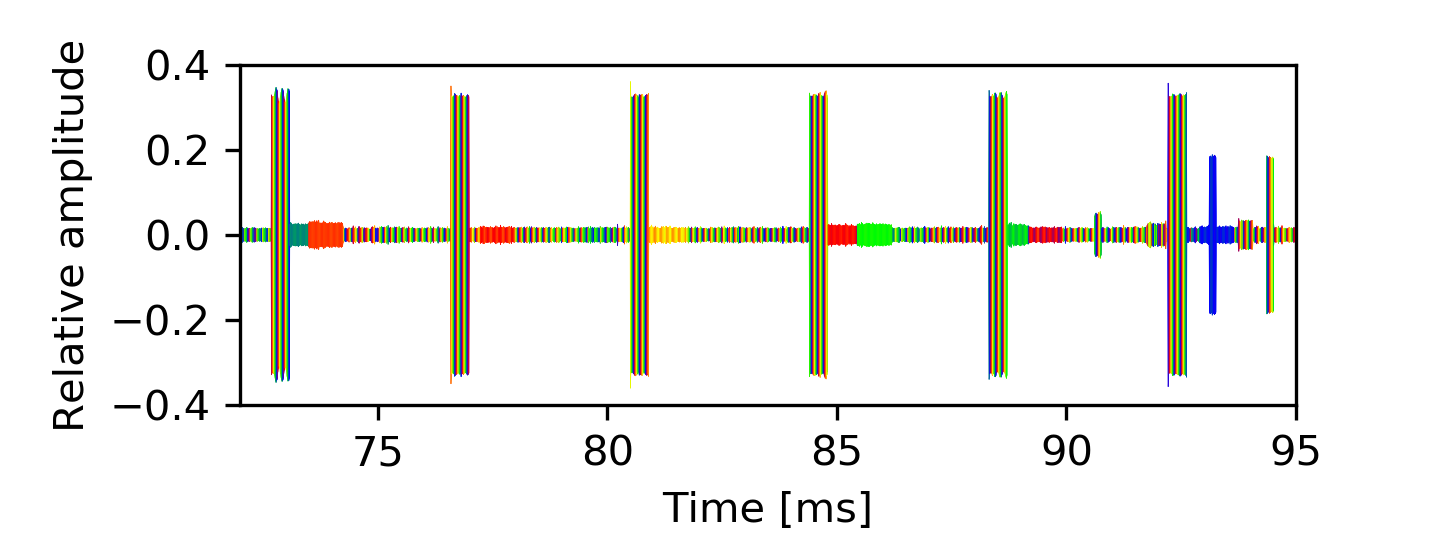}%
  \includegraphics[trim=0 2 0 8,width=.25\linewidth,clip]{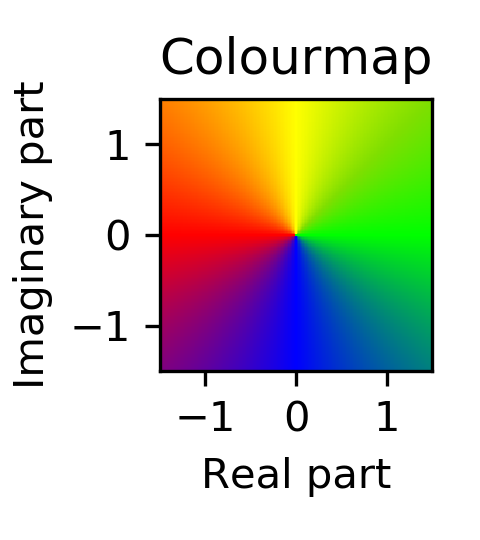}
  \caption{Signal recorded by receiver given rectangular-shaped test signal}
  \label{fig:Rect1-Ant}
\end{figure}

For visualisation of possible measurement artefacts encountered during signal transmissions with the setup introduced in Section~\ref{sec:setting}, the actual measurements recorded at the receiver while repeatedly transmitting a rectangular-shaped test signal through the line-of-sight wireless channel is depicted in \figurename~\ref{fig:Rect1-Ant}.
Here and in all subsequent plots of complex-valued signals, the amplitude of the signal is plotted both above and below the horizontal axis as in an envelope plot and the phase of the signal is represented by the colours, where green, yellow, red, and blue represent positive real, positive imaginary, negative real, and negative imaginary values, respectively.
The corresponding colourmap is provided on the right side of \figurename~\ref{fig:Rect1-Ant}.
As to be expected, common effects such as zero offsets potentially caused by offset voltages in the transmitter amplifier circuitry, carrier frequency and phase offsets, and additive white noise can be observed.
Two types of artefacts appear to be specific to the present setup:
\begin{itemize}
  \item Random-valued block-shaped bursts of similar durations showing up on a seasonal basis (i.e.\ located almost equidistantly in time) in the modulated signal (i.e.\ prior to compensation for carrier frequency offset), which suggests possible measurement artefacts on the receiver side
  \item Sporadic pulse-shaped interference, indicating the presence of signals transmitted from other devices that are sharing the considered channel
\end{itemize}
These measurement artefacts may have negative impact on channel estimation, and therefore need to be estimated and compensated for properly.

\begin{figure}[htbp]
  \centering
  \includegraphics[trim=0 2 0 8,width=\linewidth,clip]{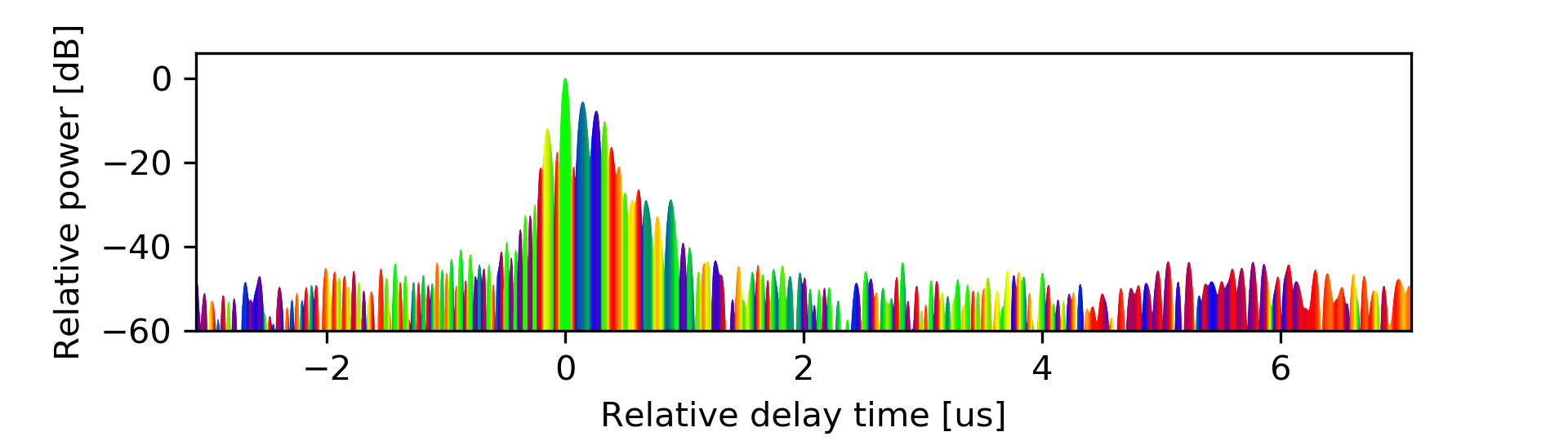}
  \caption{Raw estimate of channel impulse response}
  \label{fig:ringing}
\end{figure}

On top of the rather setting-specific measurement artefacts, another type of artefact emanating from the limitations in bandwidth and transmission time is observed while performing channel estimation in the time domain based on Fourier analysis.
This is depicted in \figurename~\ref{fig:ringing}, where an estimated channel impulse response obtained by performing the inverse Fourier transform on an estimated channel transfer function of the non-line-of-sight channel introduced in Section~\ref{sec:setting} is displayed.
Note in particular the unusual ringing tail prior to the first major impulse (at relative delay time $0$), which has no meaningful physical interpretation.
To counteract this, additional post-processing is required.

\section{Signal restoration and channel estimation}\label{sec:restore}
In this section, a signal processing routine devised for tackling the issues addressed in Section~\ref{sec:effects} is presented.
As previously mentioned, the method consists of two parts:
\begin{itemize}
  \item A signal restoration algorithm that aims to compensate for measurement artefacts
  \item A channel estimation algorithm including the design of an appropriate test signal and post-processing steps that aim to mitigate artefacts within the estimated channel impulse response
\end{itemize}
While the first part is partially specific to the hardware employed in the setup introduced in Section~\ref{sec:setting}, the second part addresses more general limitations of inferring the channel impulse response from the channel state at a limited number of frequencies.

\subsection{Signal restoration algorithm}\label{subsec:restore}
The basic idea of the restoration algorithm is to estimate and remove different offsets and artefacts step by step within an iterative procedure, initialising all intermediate steps in the first run with roughly estimated values and refining the ongoing input to each estimation with increasing number of iterations.

\begin{figure}[htbp]
  \centering
  \includegraphics[width=\linewidth]{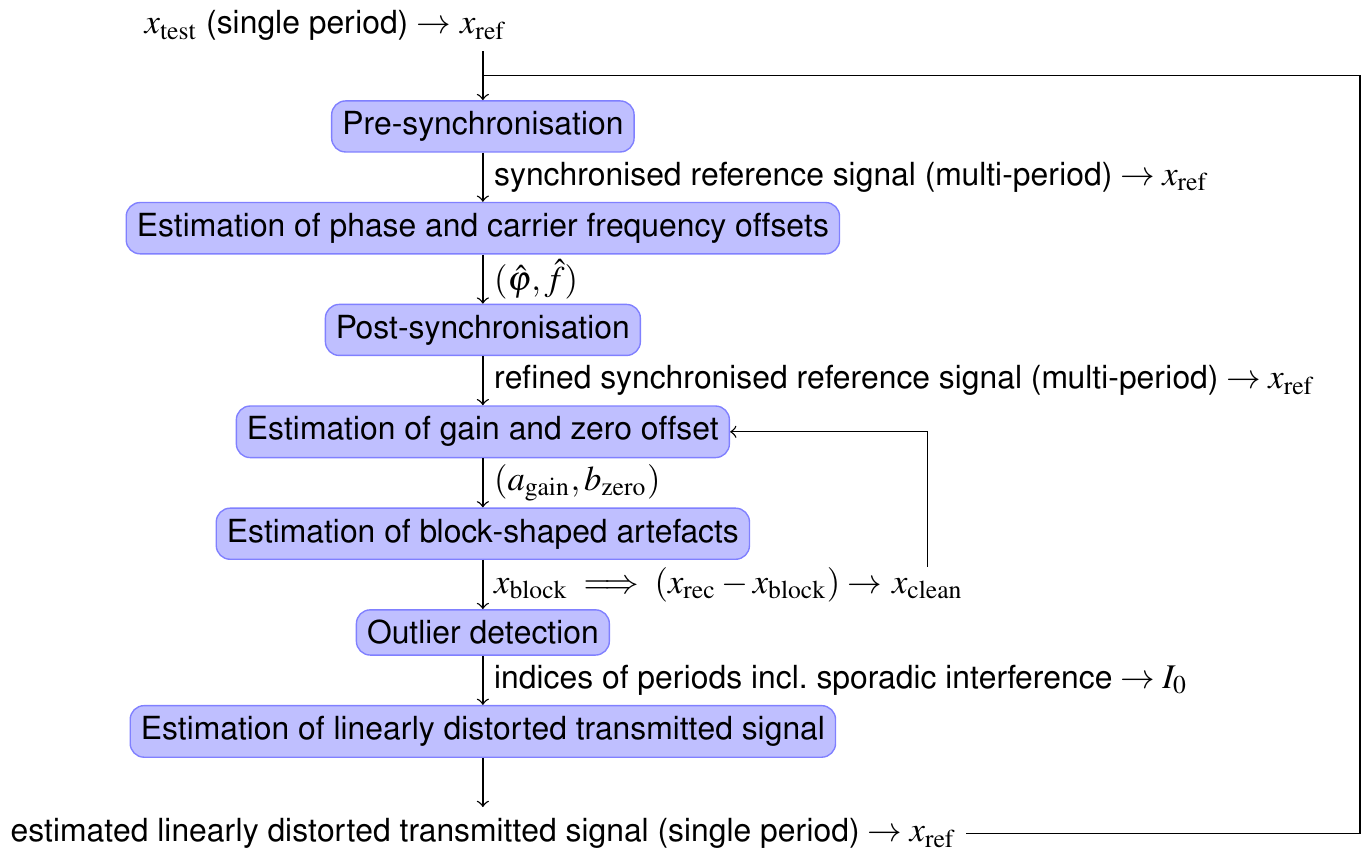}
  \caption{Block diagram of signal restoration algorithm}
  \label{fig:diagram}
\end{figure}

Henceforth, let $x_{\operatorname{test}}=\{x_{\operatorname{test}}[t]\}_{t=0}^{T-1}$ and $x_{\operatorname{rec}}=\{x_{\operatorname{rec}}[t]\}_t$ denote the one-dimensional $\mathbb{C}$-valued signals sent from the transmitter (single-period) and measured at the receiver (multi-period), respectively.
For simplicity of notation, a variable sequence, $x_{\operatorname{ref}}=\{x_{\operatorname{ref}}[t]\}_t$, is used throughout the algorithm, which represents the reference signal taken as input to the ongoing estimation and will be specified in the description of each intermediate step below.
The iterative procedure consists of the following steps conducted in succession and is illustrated in the block diagram in \figurename~\ref{fig:diagram}.

\subsubsection{Pre-synchronisation}\label{subsubsec:pre-synchr}
In general, signal synchronisation relies on detecting local peak values in the cross-correlation of the received signal $x_{\operatorname{rec}}$ against the reference signal $x_{\operatorname{ref}}$.
In the first iteration, $x_{\operatorname{ref}}$ is set to be the single-period test signal from the transmitter, $x_{\operatorname{test}}$.
In the presence of carrier frequency offset, segments in $x_{\operatorname{rec}}$ originating from the same part of $x_{\operatorname{ref}}$ exhibit similar values in amplitude but vary significantly in phase.
Therefore, only the amplitude of both $x_{\operatorname{rec}}$ and $x_{\operatorname{ref}}$ is used to compute the cross-correlation.
Taking the temporal location of each detected peak in the cross-correlation as the beginning of a period, the synchronised reference signal for the upcoming restoration steps is obtained by piecing together copies of the single-period reference signal, $x_{\operatorname{ref}}$, each padded with zeros on the right (if necessary) until the next period.

\subsubsection{Estimation of phase and carrier frequency offsets}\label{subsubsec:carr.freq.offset}
Let now $x_{\operatorname{ref}}$ be the synchronised reference signal generated according to \ref{subsubsec:pre-synchr}.
The phase and carrier frequency offsets are estimated by looking for the maximum of the average correlation of the instantaneous phase shift of the received signal $x_{\operatorname{rec}}[t]$ relative to $x_{\operatorname{ref}}[t]$ weighted by the instantaneous power, $|x_{\operatorname{ref}}[t]|^2$, against any possible instantaneous phase shift in the form of $\exp( 2\pi i f t + i\varphi )$.
More precisely, find $\hat{\varphi}$ and $\hat{f}$ such that\footnote{Here and in the sequel, $i$ refers to the imaginary unit and $\overline{x}$ refers to the complex conjugate of $x\in\mathbb{C}$.}
\begin{equation}\label{eq:carr.freq.offset}
  (\hat{\varphi},\hat{f})=\operatorname*{arg\,max}_{\varphi, f}\operatorname{Re}\sum_{t}(x_{\operatorname{rec}}[t]\mult\overline{x}_{\operatorname{ref}}[t])\cdot\overline{\exp(2\pi i f t +i\varphi)}.
\end{equation}
Note that the expression on the right hand-side of \eqref{eq:carr.freq.offset} is equal to the real part of the discrete Fourier transform ($\operatorname{FFT}$) of $x_{\operatorname{rec}}\mult\overline{x}_{\operatorname{ref}}$ evaluated at $f$ phase shifted by $-\varphi$.
Therefore, $\hat{f}$ and $\hat{\varphi}$ are determined by
\[
  \hat{f} = \operatorname*{arg\,max}_{f}\lvert\operatorname{FFT}(x_{\operatorname{rec}}\mult \overline{x}_{\operatorname{ref}})[f]\rvert,
  \,
  \hat{\varphi} = \operatorname{arg} \operatorname{FFT}(x_{\operatorname{rec}}\mult \overline{x}_{\operatorname{ref}})[\hat{f}].
\]
In order to make the estimation more precise, the resolution of the FFT is increased by padding the time-domain signal with zeros of $15$-times the original sequence length.

\subsubsection{Post-synchronisation}\label{subsubsec:post-synchr}
After the estimation of phase and carrier frequency offsets, a refined synchronised reference signal is generated with the same principle as in~\ref{subsubsec:pre-synchr}, where the peak detection is performed on the real part of the cross-correlation between the actual values (instead of only amplitude) of the demodulated received signal, $\{x_{\operatorname{rec}}[t]\cdot\exp(-2\pi i \hat{f} t -i\hat{\varphi})\}_t$, against the single-period reference signal used in~\ref{subsubsec:pre-synchr}.

\subsubsection{Estimation of gain and zero offset}\label{subsubsec:zero offsets}
Let $x_{\operatorname{clean}}$ denote a cleaned version of the received signal without measurement artefacts and let $x_{\operatorname{ref}}$ be the synchronised reference signal generated according to~\ref{subsubsec:post-synchr}.
Following the common first-order estimation, $x_{\operatorname{clean}}$ demodulated according to the results from~\ref{subsubsec:carr.freq.offset} is assumed to be of the form
\begin{equation}\label{eq:lin.regress.}
x_{\operatorname{clean}}[t]\cdot\exp(-2\pi i \hat{f} t -i\hat{\varphi}) = a_{\operatorname{gain}}\mult x_{\operatorname{ref}}[t] + b_{\operatorname{zero}} + n[t] \quad \text{f.a. }t
\end{equation}
where $a_{\operatorname{gain}}$ refers to the average value of the channel transfer function, $b_{\operatorname{zero}}$ models the constant offset in the transmitter, and $\{n[t]\}_t$ is assumed to be additive Gaussian white noise taking values in $\mathbb{C}$.
In the first iteration, $x_{\operatorname{clean}}$ is roughly approximated by the original measurements $x_{\operatorname{rec}}$.
For the linear regression model \eqref{eq:lin.regress.}, the parameters $a_{\operatorname{gain}}$ and $b_{\operatorname{zero}}$ are determined by means of the least-squares estimate
\[
\operatorname*{arg\,min}_{a,b} \sum_{t}|x_{\operatorname{clean}}\cdot\exp(-2\pi i \hat{f} t -i\hat{\varphi})-(a\mult x_{\operatorname{ref}}[t]+b)|^2.
\]

\subsubsection{Estimation of block-shaped bursts}\label{subsubsec:blocks}
Let $x_{\operatorname{art}}=\{x_{\operatorname{art}}[t]\}_t$ denote the time series of artefacts including possible noise within the received signal.
The first-order estimation model \eqref{eq:lin.regress.} leads to
\begin{equation}\label{eq:noisy blocks}
x_{\operatorname{art}}[\cdot]= x_{\operatorname{rec}}[\cdot]-(a_{\operatorname{gain}}\mult x_{\operatorname{ref}}[\cdot] + b_{\operatorname{zero}})\cdot\exp(2\pi i \hat{f} \cdot[\cdot] +i\hat{\varphi}).
\end{equation}
As stated in Section~\ref{sec:effects}, $x_{\operatorname{art}}$ mainly consists of block-shaped bursts occurring randomly on a seasonal basis (i.e.\ almost equidistantly in time).
The temporal locations of those blocks are first detected by means of local peak detection in the seasonal difference of a low-pass filtered version of $x_{\operatorname{art}}$, where the season duration and width of the filter kernel are initialised with a rough estimation of the block length based on visual inspection and refined with increasing number of iterations.
The first detected locations of blocks are then processed through a selection algorithm that removes excessive points or adds extra points in between so as to make potential block boundaries more evenly distributed over time in accordance with the observations in Section~\ref{sec:effects}.
Let $x_{\operatorname{blocks}}=\{x_{\operatorname{blocks}}[t]\}_t$ denote the time series of estimated block-shaped artefacts.
Since most of the visible blocks exhibit linear dependency on time, the value of $x_{\operatorname{blocks}}$ is determined through piece-wise linear regression.

\subsubsection{Outlier detection and estimation of linear effect}\label{subsubsec:linear effect}
In contrast to the seasonal block-shaped bursts, which are inevitable during the estimation of the linearly distorted signal and therefore need to be estimated and removed from the received signal, the other type of measurement artefacts addressed in Section~\ref{sec:effects}, namely the sporadic pulse-shaped interference, can be disposed of through outlier detection.
That is, the linearly distorted transmitted signal directly resulting from the channel impact is estimated through
\begin{multline}\label{eq:linear effect}
  \mathbf{1}_{[0,T\mathclose{[}}[\cdot]\mult\operatorname*{mean}_{\iota\in I\setminus I_0}\{(x_{\operatorname{rec}}[\iota+{\cdot}]-x_{\operatorname{blocks}}[\iota+{\cdot}])\\\cdot\exp(-2\pi i \hat{f} \cdot[\iota+{\cdot}] -i\hat{\varphi}_{\iota})\}-b_{\operatorname{zero}}
\end{multline}
where $I$ refers to the index set of period beginnings in the synchronised reference signal generated according to~\ref{subsubsec:post-synchr}, $I_0$ refers to the collection of period indices associated with the detected outliers, and $\hat{\varphi}_{\iota} = \hat{\varphi} + \Delta \varphi_{\iota}$ with $\Delta \varphi_{\iota}$ referring to the final phase correction towards $x_{\operatorname{test}}$ for period $\iota$.
For the outlier detection, the energy of each summand in \eqref{eq:linear effect} for all $\iota\in I$ is used for reference and periods with reference value higher than $3/2$ the median of the empirical distribution function are discarded.

\subsubsection*{Iterative procedure}
After the first run, steps \ref{subsubsec:pre-synchr} through \ref{subsubsec:linear effect} are carried out iteratively, with $x_{\operatorname{ref}}$ used in step~\ref{subsubsec:pre-synchr} updated to the latest output of step~\ref{subsubsec:linear effect} and $x_{\operatorname{clean}}$ used in step~\ref{subsubsec:zero offsets} updated to $x_{\operatorname{rec}}-x_{\operatorname{blocks}}$ with the latest estimation of $x_{\operatorname{blocks}}$ from step~\ref{subsubsec:blocks}.

\subsection{Channel estimation}\label{subsec:channel estimation}
The channel estimation is performed by repeatedly transmitting a pre-defined test signal through the channel, recording the received signal, applying the signal restoration algorithm from Section~\ref{subsec:restore} to the received signal, estimating the channel transfer function from the restored signal by means of the discrete Fourier transform, and inferring the channel impulse response from the estimated transfer function through a post-processing algorithm based on Fourier analysis.

\subsubsection{Test signal and frequency domain channel estimation}\label{subsubsec:channel transfer function}
The waveform used in the test signal for channel estimation is based on a Zadoff-Chu sequence of prime length, cf.~\cite{Chu72}.
In general, Zadoff-Chu sequences of prime length are a class of signals particularly well-suited for channel estimation due to their zero-autocorrelation property, which makes them ideal for synchronisation, and their uniformity of amplitude in both the time and frequency domains, which results in minimal crest factor.
In particular, their FFTs are always bounded away from zero.
In this setting, the test signal $x_{\operatorname{test}} = \{ x_{\operatorname{test}}[t] \}_{t = 0}^{T - 1}$ used for channel estimation consists of 4 repetitions of an arbitrary Zadoff-Chu sequence of prime length $N_{\operatorname{ZC}}$, denoted by $x_{\operatorname{ZC}} = \{ x_{\operatorname{ZC}}[t] \}_{t = 0}^{N_{\operatorname{ZC}} - 1}$, i.e.,
\[
  x_{\operatorname{test}}[t] =
  \begin{cases}
       x_{\operatorname{ZC}}[t \bmod N_{\operatorname{ZC}}] & \text{if $t < 4 \mult N_{\operatorname{ZC}}$,}
    \\ 0                                                    & \text{if $4 \mult N_{\operatorname{ZC}} \le t < T$.}
  \end{cases}
\]
In order to prevent the channel estimation from being affected by potential boundary effects caused by the fading channel, the sequence length $N_{\operatorname{ZC}}$ is chosen such that the actual temporal length of $x_{\operatorname{ZC}}$ is at least twice the expected maximum delay time of the channel impulse response, so that only the first and last half-repetitions within $x_{\operatorname{test}}$ may be corrupted by boundary effects and the remaining part in the received signal is cyclic.
On the receiver side, the recorded signal is processed through the signal restoration procedure presented in Section~\ref{subsec:restore}.
From the final output of the algorithm, the potentially corrupted parts originating from the first and last half-repetitions of $x_{\operatorname{ZC}}$ are removed, and the remaining 3 repetitions are averaged, resulting in a signal $x_{\operatorname{est}} = \{ x_{\operatorname{est}}[t] \}_{t = 0}^{N_{\operatorname{ZC}} - 1}$.
Both $x_{\operatorname{ZC}}$ and $x_{\operatorname{est}}$ can now be regarded as sampled versions of band-limited periodic signals with the same period length, which allows the use of the discrete Fourier transform (FFT) to compute their frequency domain representations.
This leads to an estimate of the channel transfer function sampled at a finite number of points within the band centered at the carrier frequency, which is denoted by $\hat{H} = \{ \hat{H}[f] \}_{f = 0}^{N_{\operatorname{ZC}} - 1}$, with
\begin{equation}
    \hat{H}[f]
  = \frac{\operatorname{FFT}(x_{\operatorname{est}})[f]}{\operatorname{FFT}(x_{\operatorname{ZC}})[f]}
  \quad \text{for $f = 0, \ldots, N_{\operatorname{ZC}} - 1$}
  \label{eq:channel transfer function}
\end{equation}
where the first and second halves of the frequency indices correspond to the actual frequencies from the intervals that are half the sampling rate in width and located to the right and left of the carrier frequency, respectively.
Note that, due to the dilation property of the FFT, the resolution of $\hat{H}$ is proportional to $N_{\operatorname{ZC}}$ and thus tunable in general.

\subsubsection{Time domain channel estimation}\label{subsubsec:impulse response}
Let $H \colon \mathbb{R} \longrightarrow \mathbb{C}$ and $h \colon \mathbb{R} \longrightarrow \mathbb{C}$ denote the actual channel transfer function and the channel impulse response, respectively, which are by their nature both continuous and aperiodic, and let $\mathcal{F}$ denote the continuous Fourier transform.
It holds that
\begin{equation}
    h
  = \mathcal{F}^{-1} H
  ,
  \label{eq:ift}
\end{equation}
which provides a theoretical basis for inferring the channel impulse response from the channel state information in the frequency domain.
In a practical discrete-time setting, however, estimating $h$ by simply replacing $H$ in~\eqref{eq:ift} by the discrete estimate $\hat{H}$ according to~\eqref{eq:channel transfer function} and performing the inverse discrete Fourier transform (IFFT) instead of $\mathcal{F}^{-1}$ would lead to artefacts which render the estimation result difficult to interpret as illustrated in \figurename~\ref{fig:ringing}.
In general, due to the dilation property of the Fourier transform, when operating in a discrete (periodic) setting and switching between the time and frequency domains, the sequence length in the one domain is proportional to the resolution in the other.
Therefore, problems while inferring the continuous aperiodic function $h$ from the discrete finite sequence $\hat{H} = \{ \hat{H}[f] \}_{f = 0}^{N_{\operatorname{ZC}} - 1}$ instead of the continuous aperiodic function $H$ are two-fold:
\begin{itemize}
  \item
    Sampling in the frequency domain results in a periodic time domain, which may cause aliasing in general.
    However, the period length, i.e., the maximum possible delay time that can be captured by the estimated channel impulse response, is equal to the reciprocal of the sample spacing within $\hat{H}$, which is exactly the temporal length of the sequence $x_{\operatorname{ZC}}$ (${} = N_{\operatorname{ZC}} / (\text{sampling rate of $x_{\operatorname{ZC}}$})$).
    Thus, the choice of $N_{\operatorname{ZC}}$ specified in~\ref{subsubsec:channel transfer function} excludes the presence of aliasing in the estimated channel impulse response.
  \item
    Information loss in the frequency domain caused by band restriction results in sacrificing the resolution in the time domain.
    On the one hand, this effect can be compensated for by extending $\hat{H}$ through zero-padding.
    On the other hand, zero-padding in the frequency domain is equivalent to multiplication by a rectangular window and thus corresponds to convolution with a $\operatorname{sinc}$ kernel in the time domain; for $\hat{H}$ not vanishing at the boundaries, this may introduce extra $\operatorname{sinc}$-shaped tails around each impulse within the estimated channel impulse response.\footnote{Note that, even without zero-padding, restricting the support in the frequency domain to a finite band causes discontinuity in the corresponding periodic signal used in the IFFT, as long as the values at the left and right boundary of the band differ from one another.  This, in general, results in possible $\operatorname{sinc}$-shaped ringing in the time domain.}
    Therefore, in the present approach, $\hat{H}$ is first multiplied by a Dolph-Chebyshev window~\cite{Dol46} (centred around the carrier frequency), $w$, with attenuation set to \SI{60}{dB}\footnote{Note that, as a side effect of windowing the frequency domain representation, the time domain is low pass filtered, which results in some loss of sharpness in the estimated channel impulse response.  Choosing the Dolph-Chebyshev window ensures maximal possible sharpness for a given attenuation of tails.  The attenuation is set to $\SI{60}{dB}$, as, at this level, the residual tails are already below the measured noise floor.}, and then extended to $16$-times the original length through zero-padding ($\operatorname{pad}_{0}^{15}$).
\end{itemize}
Overall, the channel impulse response is estimated through
\[
  \operatorname{IFFT} \biggl( \operatorname{pad}_{0}^{15} \biggl( \frac{\operatorname{FFT}(x_{\operatorname{est}})}{\operatorname{FFT}(x_{\operatorname{ZC}})} \cdot w \biggr) \biggr)
  .
\]
\mbox{}

\section{Results}\label{sec:results}
In this section, partial and overall effects of the signal restoration algorithm and channel estimation procedure proposed in Section~\ref{sec:restore} are evaluated on real measurements acquired using the setup from Section~\ref{sec:setting}.
For each evaluation, the signal restoration algorithm is run for $10$ iterations.

\subsection{Evaluation of signal restoration}
Since the core component of the measurement artefacts consists in block-shaped bursts specific to the receiving equipment, a simple rectangular-shaped test signal is transmitted through the cable-based setup given in Section~\ref{sec:setting} for evaluating restoration steps~\ref{subsubsec:zero offsets} and~\ref{subsubsec:blocks}.

\begin{figure}[htbp]
  \centering
  \includegraphics[trim=0 8 0 40,width=\linewidth,clip]{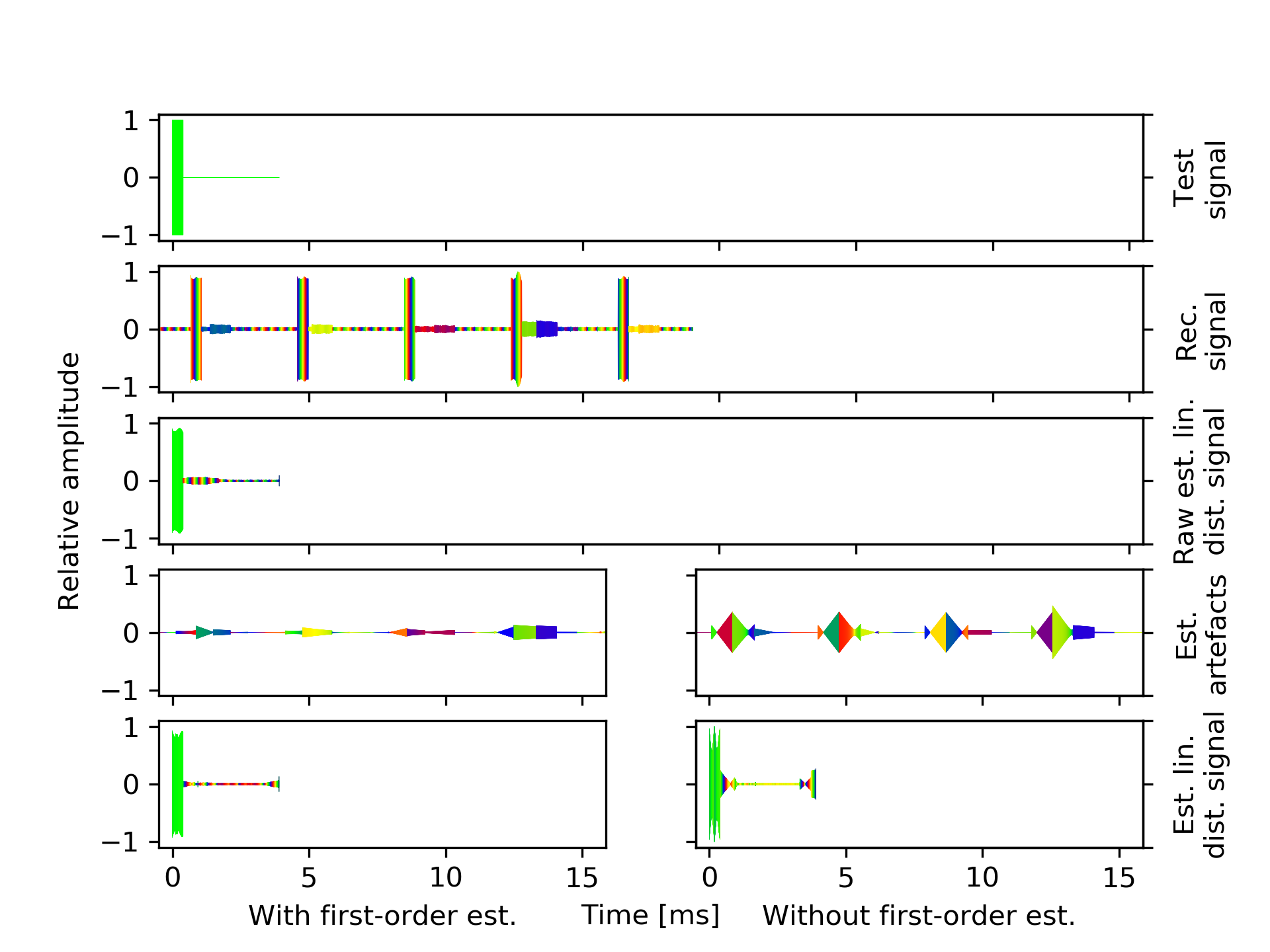}
  \caption{Effect of signal restoration}
  \label{fig:effect-blocks}
\end{figure}

In \figurename~\ref{fig:effect-blocks}, following plots are generated for comparison:
\begin{itemize}
  \item Test signal from the transmitter, $x_{\operatorname{test}}$
  \item Original measurements recorded at the receiver, $x_{\operatorname{rec}}$
  \item Estimated linearly distorted transmitted signal without employing restoration step~\ref{subsubsec:blocks}
  \item Estimated artefacts output from step~\ref{subsubsec:blocks}, $x_{\operatorname{blocks}}$, with vs.\ without step~\ref{subsubsec:zero offsets}
  \item Estimated linearly distorted transmitted signal, employing restoration step~\ref{subsubsec:blocks}, with vs.\ without step~\ref{subsubsec:zero offsets}
\end{itemize}
Here, among the subplots in the last two rows, the ones on the left are generated exactly as proposed in the restoration algorithm, whereas the ones on the right are generated by setting $x_{\operatorname{art}}$ used in~\ref{subsubsec:blocks} to $x_{\operatorname{rec}}$ instead of the expression on the right-hand side of~\eqref{eq:noisy blocks}.
When examining the plots on the left, the similarity of $x_{\operatorname{blocks}}$ to the artefacts visible within $x_{\operatorname{rec}}$ indicates that most of the block-shaped bursts are accurately estimated and removed from the original measurements through the signal restoration algorithm.
This is further illustrated in the estimated linearly distorted transmitted signal depicted at the bottom left corner, considering the well-preserved step-shape and the rapid decay as opposed to the particularly slow decay of the raw estimate depicted in the third row.
In addition, when examining the plots on the right, the inaccurate estimate of the measurement artefacts and the significantly distorted step-shape in the corresponding estimated linearly distorted transmitted signal highlight the necessity of the first-order estimation in step~\ref{subsubsec:zero offsets} within the algorithm.

\subsection{Impact on channel estimation}

\begin{figure}[htbp]
  \centering
  \includegraphics[trim=0 8 0 40,width=\linewidth,clip]{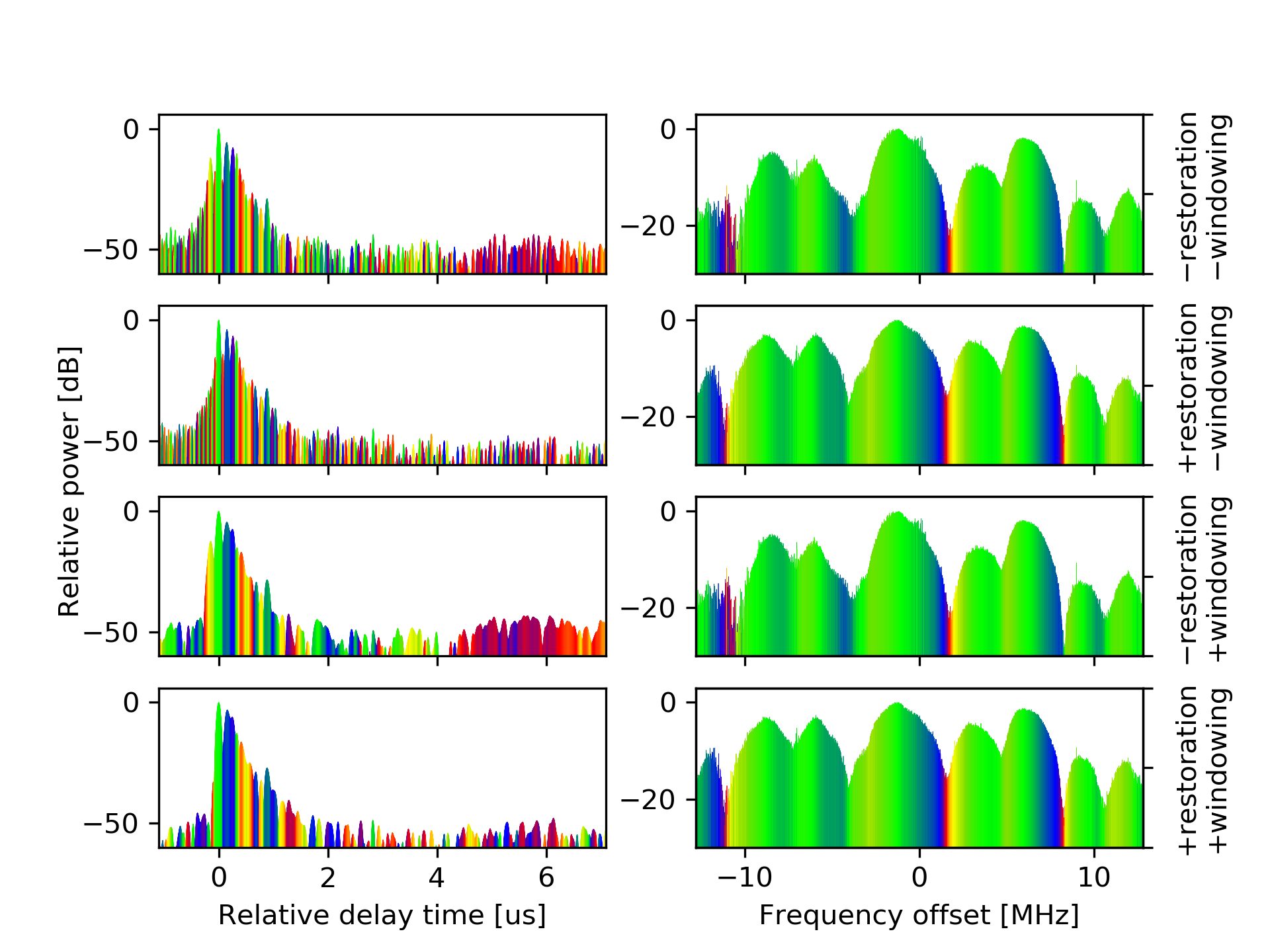}
  \caption{Effect of proposed algorithm on channel estimation}
  \label{fig:channel-est}
\end{figure}

For evaluating the effects of both the signal restoration algorithm and the post-processing steps, the test signal with the repeating Zadoff-Chu sequence specified in~\ref{subsubsec:channel transfer function} is transmitted through the non-line-of-sight channel configured according to Section~\ref{sec:setting}.

\figurename~\ref{fig:channel-est} provides a visualisation of the channel estimation in both the time and frequency domains performed in the following order:
\begin{enumerate}
  \item Only conducting steps \ref{subsubsec:pre-synchr} through \ref{subsubsec:post-synchr} for basic synchronisation and skipping the windowing step in the frequency domain from~\ref{subsubsec:impulse response}
  \item Conducting all restoration steps \ref{subsubsec:pre-synchr} through \ref{subsubsec:linear effect} but skipping the windowing step in the frequency domain from~\ref{subsubsec:impulse response}\label{item:no-window}
  \item Only conducting steps \ref{subsubsec:pre-synchr} through \ref{subsubsec:post-synchr} for basic synchronisation but without resolution of measurement artefacts, while keeping all post-processing steps in~\ref{subsec:channel estimation}\label{item:post-synchr}
  \item Keeping all steps as proposed in \ref{subsec:restore}~and~\ref{subsec:channel estimation}\label{item:restored}
\end{enumerate}
Starting with the baseline approach, the result of channel estimation without restoration or post-processing steps illustrated in the first row exhibits significant disturbance prior to the first dominant impulse (at relative delay time $0$) along with a high noise floor in the time domain and visible irregularities in the frequency domain.
In the absence of the windowing step in the frequency domain, the inferred channel impulse response displayed in the second row contains the same unnatural pre-ringing as the one showing up in \figurename~\ref{fig:ringing} as discussed in Section~\ref{sec:effects}, which obscures the shape and extent of the actual power delay profile.
In contrast, this pre-ringing is significantly mitigated in the time domain of the third and fourth rows, which demonstrates the efficacy of the frequency-domain windowing procedure proposed in~\ref{subsubsec:impulse response}.
Among the remaining estimates in the last two rows, three major differences are worth noting:
First, in the frequency domain, the channel transfer function inferred from the fully restored signal appears smoother, particularly in the higher frequency region (towards boundaries of plot).
Similarly, in the time domain, the full restoration algorithm also contributes to a lower noise floor in the estimated channel impulse response.
Moreover, while a pre-echo is only visible in the time-domain representation of the example displayed in the third row (red to yellow section prior to relative delay time $0$), this effect cannot be reproduced when evaluating the underlying estimation method with other comparable measurements acquired at the same physical location, which indicates an estimation error caused by measurement artefacts that would have been resolved through the missing restoration steps \ref{subsubsec:zero offsets} through \ref{subsubsec:linear effect}.
Despite the lack of ground truth knowledge in the employed setup, the comparison in the above three aspects suggests that the proposed signal restoration algorithm for mitigating the setting-specific measurement artefacts indeed aids in improving the quality of channel estimation.

\subsection{Convergence of iterative procedure}
For evaluating the convergence behaviour of the proposed iterative signal restoration routine, the energy of the difference between the outputs of step~\ref{subsubsec:linear effect} from consecutive iterations is calculated for the Zadoff-Chu test signal as introduced in~\ref{subsubsec:channel transfer function} and presented in \tablename~\ref{tab:conv}.
It turns out that the estimated linearly distorted transmitted signal converges from the 3\textsuperscript{rd} iteration, which confirms the stability of the proposed algorithm.
\begin{table}[htbp]
  \caption{Convergence of iterative signal restoration routine}
  \label{tab:conv}
  \(
    \displaystyle
    \hfill
    \begin{array}{rl@{\qquad}rl}
      \text{Iteration} & \text{Energy difference} & \text{Iteration} & \text{Energy difference}
      \\  1 & 0.0040479            &  6 & 2.5285 \cdot 10^{-7}
      \\  2 & 0.00055984           &  7 & 2.4055 \cdot 10^{-7}
      \\  3 & 2.8392 \cdot 10^{-7} &  8 & 2.2315 \cdot 10^{-7}
      \\  4 & 3.9208 \cdot 10^{-7} &  9 & 2.6561 \cdot 10^{-7}
      \\  5 & 3.3383 \cdot 10^{-7} & 10 & 2.4544 \cdot 10^{-7}
    \end{array}
    \hfill
  \)
\end{table}

\section{Conclusion}\label{sec:conclusion}
In this work, a channel estimation procedure including signal restoration and post-processing is designed for a concrete channel sounding setup based on SDRs.
Experimental results confirm that the proposed algorithm is able to compensate for the shortcomings of the hardware and improve the quality of the subsequent channel estimation, which in particular suggests the potential of the considered setup being a cost-effective yet reasonable alternative to classical channel sounding equipment.
In future research, the method is planned to be employed in an SDR-based channel measurement campaign.

\bstctlcite{IEEEexample:BSTcontrol}
\bibliographystyle{IEEEtran}
\bibliography{IEEEabrv,bib}

\end{document}